\documentstyle[preprint,aps]{revtex}
\begin{document}
\draft
\title{Dynamic Scaling of Ion--Sputtered Surfaces}

\author{R. Cuerno$^{\dag}$ and A.-L. Barab\'asi$^{\dag}$ $^{\ddag}$}

\address{$^{\dag}$ Center for Polymer Studies and Dept. of Physics,
 Boston University, Boston, MA 02215 \\
$^{\ddag}$ IBM T.\ J.\ Watson Research Center, P.O. Box 218,
Yorktown Heigths, NY 10598}

\date{\today}

\maketitle

\begin{abstract}

We derive a stochastic nonlinear equation to describe the evolution and
scaling properties of surfaces eroded by ion bombardment.
The coefficients appearing in the equation can be calculated
explicitly in terms of the physical parameters characterizing
the sputtering process.  We find that transitions may take place
between various scaling behaviors when experimental parameters such as the
angle of incidence of the incoming ions or their average penetration depth,
are varied.
\end{abstract}

\pacs{PACS numbers: 79.20.Rf, 64.60.Ht, 68.35.Rh}





Recently there has been much interest in understanding the formation
and roughening of nonequilibrium interfaces. Applications range from
fluid flow in porous media to bacterial colonies or molecular beam
epitaxy \cite{revrough}. A common feature of most rough interfaces
observed experimentally or in discrete models is that their roughening
follows simple scaling laws.  The associated scaling exponents can be
obtained using numerical simulations or stochastic evolution
equations.
The morphology and dynamics of a rough interface can be
characterized by the surface width, $w(t,L)$, that scales as
$w^2(t,L)=\langle [h(\mbox{\boldmath $r$},t)-{\overline h(t)}]^2
\rangle = L^{2 \alpha} f(t/L^{z})$, where $\alpha$ is the roughness
exponent for the interface $h(\mbox{\boldmath $r$},t)$ and the dynamic
exponent $z$ describes the scaling of the relaxation times with the
system size $L$; $\overline h(t)$ is the mean height of the interface
at time $t$ and $\langle \;\; \rangle$ denotes both ensemble and
space average. The scaling function $f$ has the properties $f(u \to 0)
\sim u^{2\alpha/z}$ and $f(u \to \infty) \sim$ const.

Much of the attention has focused so far on the kinetics of interfaces
generated in {\em growth} processes. However, for a class of
technologically important phenomena, such as sputter etching, the
surface morphology evolves as a result of {\em erosion} processes
\cite{revsput}.  Motivated by the advances in understanding growth,
recently a number of experimental studies have
focused on the scaling properties of surfaces eroded by ion
bombardment \cite{eklund,krim,chason}.  For graphite bombarded with 5
keV Ar ions, Eklund {\em et al.}  \cite{eklund} reported $\alpha
\simeq 0.2-0.4$, and $z \simeq 1.6-1.8$, values consistent with the
predictions of the Kardar-Parisi-Zhang (KPZ) equation in 2+1 dimensions
\cite{kpz,numer}. Similarly Krim {\em et al.}  \cite{krim} observed a
self--affine surface generated by 5 keV Ar bombardment of an Fe
sample, with a larger exponent, $\alpha = 0.52$. On the other hand,
there exists ample evidence about the development of a periodic ripple
structure in sputter etched surfaces (see e.g.\ \cite{carter}).
Chason {\em et al.} \cite{chason} have recently
studied the dynamics of such eroded surfaces for both SiO$_2$ and Ge
bombarded with Xe ions at 1 keV, and found that it differs from the dynamics
expected for the self-affine morphologies observed in \cite{eklund}
and \cite{krim}.

In this paper we investigate the large scale properties of
ion-sputtered surfaces aiming to understand in an unified framework
the various dynamics and scaling behaviors of the experimentally observed
surfaces.  For this we derive a stochastic nonlinear equation that
describes the time evolution of the surface height.  The coefficients
appearing in the equation are functions of the physical parameters
characterizing the sputtering process.  We find that transitions may
take place between various surface morphologies as the experimental
parameters (e.g.\ angle of incidence, penetration depth) are varied.
Namely, at short length-scales the equation describes the development
of a periodic ripple structure, while at larger length-scales the
surface morphology may be either logarithmically ($\alpha=0$) or
algebraically ($\alpha > 0$) rough.  Usually stochastic equations
describing growth models are constructed using symmetry principles and
conservation laws.  In most cases it is difficult to derive the
equation from a given discrete model. In contrast, here we show that
for sputter eroded surfaces the growth equation can be derived
directly from a simple model of the elementary processes taking place
in the system.

Most of the theoretical approaches focusing on the scaling properties
of sputter roughened surfaces have assumed that essentially all
relevant processes take place at the surface, and that nonlinear
effects would appear only due to non--local effects such as shadowing
\cite{bales}.  However, ion-sputtering is in
general determined by atomic processes taking place along a finite
penetration depth inside the bombarded material. The incoming ions
penetrate the surface and transfer  their kinetic energy to the atoms of the
substrate by inducing
cascades of collisions among the substrate atoms, or through other
processes such as electronic excitations. Whereas most of the atoms which
are actually sputtered are those located at the surface,
the scattering events that
might lead to sputtering take place within a certain layer of average
depth $a$. As described by Sigmund's transport theory of sputtering
\cite{sig}, the
average value of the ion deposition depth  depends on the
energy of the bombarding ions, their angle of incidence, the
microscopic structure of the target material and the features of the
scattering processes taking place inside the sample.

A convenient picture of the ion bombardment process is sketched in the
inset of Fig.\ \ref{fig1}.  According to it the ions penetrate a
distance $a$ inside the solid before they completely spread out their
kinetic energy with some assumed spatial distribution.  An ion
releasing its energy at point $P$ in the solid contributes an amount
of energy to the surface point $O$, that may induce the atoms in $O$
to break their bonds and leave the surface.  Following \cite{sig,bh},
we consider that the average energy deposited at point $O$ due to the
ion arriving at $P$ follows the Gaussian distribution
\begin{equation}
E_D(\mbox{\boldmath $r'$})  =  \frac{\epsilon}{(2\pi)^{3/2} \sigma
\mu^2} \exp \left\{ - \frac{z'^2}{2 \sigma^2} -
\frac{x'^2+y'^2}{2 \mu^2} \right\}.
\label{gaussian}
\end{equation}
In (\ref{gaussian}) $z'$ is the distance measured along the ion
trajectory, and $x'$, $y'$ are measured in the plane perpendicular to
it (see Fig.\ \ref{fig1}; for simplicity in the figure $x'$ has been
set to 0); $\epsilon$ denotes the total energy carried
by the ion, $a$ is the average penetration length and $\sigma$ and
$\mu$ are the widths of the distribution in directions parallel and
perpendicular to the incoming beam respectively.
However, the sample is subject to an uniform flux $J$ of
bombarding ions. A large number of ions penetrate
the solid at different points simultaneously and the velocity of
erosion at $O$ depends directly on the total
power ${\cal E}_O$ contributed by all the ions deposited within
the range of the distribution (\ref{gaussian}).  If we ignore
shadowing effects among neighboring points, as well as further
redeposition of the eroded material, the velocity of erosion at $O$ is
given by
\begin{equation}
v = p \; \int_{{\cal R}} d\mbox{\boldmath $r$} \; \Phi(\mbox{\boldmath $r$})
\; E_D(\mbox{\boldmath $r$}),
\label{vel}
\end{equation}
where the integral is taken over the region ${\cal R}$ of all the points
at which the deposited energy contributes to ${\cal E}_O$,
$\Phi(\mbox{\boldmath $r$})$
is a local correction to the uniform flux $J$ and $p$ is a proportionality
constant between power deposition and rate of erosion.
In the following we outline the basic steps in the calculation
of $v$; further details can be found in Refs.\ \cite{bh,us}.

The calculation of (\ref{vel}) is most efficiently perfomed in the
{\em local} coordinate system $(X,Y,Z)$ shown in Fig.\ \ref{fig1}.
The ion beam lies
in the $X$-$Z$ plane, forming an angle $\varphi$ with the $Z$ axis,
$Z$ being normal to the surface at $O$.  To simplify the calculations,
we assume that: (a) the radii of curvature ($R_X$,
$R_Y$) of the surface at $O$ are much larger than the penetration
depth $a$, so that only terms up to first order in $a/R_{X,Y}$ are
kept; (b) the curvatures attain their maximum and minimum values along
either of the $X$ and $Y$ directions, and thus we can expand the value
of the surface height at $O$ by taking a second order approximation in
the ($X$,$Y$) coordinates and consistently ignoring cross term
contributions.  Performing the integral (\ref{vel})
the velocity $v(\varphi,R_X,R_Y)$ is found to be a
function of the angle $\varphi$ and the curvatures $1/R_{X,Y}$ \cite{bh}.

Finally, from the expression of $v$ we can obtain the equation of
motion for the profile. Now it is convenient to use the {\em laboratory}
coordinate frame $(x,y,h)$. In the absence of overhangs the surface can be
described by a single valued height function $h(x,y,t)$, measured from an
initial flat configuration which is taken to lie in the ($x$,$y$)
plane.  The ion beam is parallel to the $x$-$h$ plane forming an angle
$0< \theta < \pi/2$ with the $z$ axis.  The time evolution of $h$ is
given by
\begin{equation}
\frac{\partial h(x,y,t)}{\partial t} \simeq -
v(\varphi,R_X,R_Y),
\label{hv}
\end{equation}
where $\varphi$ is the angle of the beam direction with the
local normal to the surface at $h(x,y)$. Now $\varphi$
is a function of the angle of incidence $\theta$ and the values of the
local slopes $\partial_x h$ and $\partial_y h$, and can be expanded in
powers of the latter. We will assume that the surface varies
smoothly enough so that products of derivatives of $h$ can be neglected
for third or higher orders.

At this stage additional
relevant physical processes must be taken into account to describe the
evolution of the surface. First, the bombarding ions reach the surface at
random positions and times. We account for the stochastic arrival of
ions by adding to (\ref{hv}) a
Gaussian white noise $\eta(x,y,t)$ with zero mean and variance
proportional to the flux $J$.  Second, at finite temperature atoms
diffuse on the surface \cite{eklund,chason}.  To include this surface
self-diffusion we allow for a term $-K \nabla^2 (\nabla^2 h)$
\cite{herr}, where $K$ is a temperature dependent positive
coefficient.  Expanding (\ref{hv}) and
adding the noise and the surface-diffusion terms we
obtain the equation of motion \cite{notexi}
\begin{equation}
\frac{\partial h}{\partial t} = -v_0 + \gamma
\frac{\partial h}{\partial x} + \nu_x \frac{\partial^2 h}{\partial x^2} +
\nu_y \frac{\partial^2 h}{\partial y^2} +
\frac{\lambda_x}{2} \left(
\frac{\partial h}{\partial x} \right)^2 + \frac{\lambda_y}{2} \left(
\frac{\partial h}{\partial y} \right)^2 - K \nabla^2 (\nabla^2 h) + \eta .
\label{eqn}
\end{equation}
{}From (\ref{hv}) we can compute the expressions for the coefficients
appearing in (\ref{eqn}) in terms of the physical parameters
characterizing the sputtering process.  To simplify the discussion we
restrict ourselves to the symmetric case $\sigma=\mu$.
The general case is discussed in \cite{us}.  If we write $F \equiv
(\epsilon J p/\sqrt{2\pi}) \exp(-a_{\sigma}^2/2)$, $s \equiv \sin
\theta$, $c \equiv \cos \theta$ and $a_{\sigma} \equiv a/\sigma$, we find for
the
coefficients in (\ref{eqn})
\begin{eqnarray}
v_0 & = & \frac{F}{\sigma} c \;\;,\;\; \gamma = \frac{F}{\sigma} s
(a_{\sigma}^2 c^2 - 1) , \nonumber \\
\lambda_x & = & \frac{F}{\sigma} c \left\{ a_{\sigma}^2
(3s^2-c^2) - a_{\sigma}^4 s^2 c^2 +1 \right\} , \nonumber \\
\lambda_y & = & -\gamma \frac{c}{s} , \label{coefs2} \\
\nu_x & = & \frac{F}{2} a_{\sigma} \left\{ 2 s^2- c^2 -
a_{\sigma}^2 s^2 c^2 \right\} , \nonumber \\
\nu_y & = & -\frac{F}{2} a_{\sigma} c^2 . \nonumber
\end{eqnarray}
Consistent with the direction of the bombarding beam and the choice
of coordinates made, the terms in (\ref{eqn}) are symmetric under $y
\rightarrow -y$ but not under $x \rightarrow -x$, while for $\theta
\rightarrow 0$ we get $\gamma = \xi_x = \xi_y = 0$, $\lambda_x =
\lambda_y$ and $\nu_x = \nu_y$ \cite{note1}.  The
equation studied in Ref.\ \cite{bh} corresponds to the deterministic
linear version of (\ref{eqn}), i.\ e.\
$\lambda_x=\lambda_y=\eta=0$.

If $\nu_x$ and $\nu_y$ are positive, the surface diffusion term is
expected to contribute negligibly as a relevant surface relaxation
mechanism when we probe the system at increasingly large length
scales. Scaling properties are then described by the anisotropic KPZ
equation (AKPZ), which predicts two possible behaviors depending on
the relative signs of the coefficients
$\lambda_x$ and $\lambda_y$ \cite{wolf,bruins}.  If $\lambda_x
\lambda_y > 0$, then $\alpha=0.38$ and $z=1.6$, the surface
width $w(L,t)$ increases algebraically, being characterized by the
exponents of the KPZ equation in 2+1 dimensions \cite{numer}. For $\lambda_x
\lambda_y < 0$, the nonlinear terms $\lambda_x$ and $\lambda_y$
become irrelevant, and the width grows only logarithmically, i.e.
$\alpha=0$.

In our case $\nu_y$ is always negative, while $\nu_x$ can
change sign as $\theta$ and $a_{\sigma}$ are varied. A negative
surface tension generates an instability in the system.  The same kind
of instability is known to take place in the Kuramoto--Sivashinsky
(KS) equation \cite{ks}, which is the {\em noisyless} and isotropic
version of (\ref{eqn}). It has been argued for the KS equation that in
1+1 dimensions $\nu$ renormalizes to a positive value \cite{ks1+1}, and
the large length scale behavior is described by the KPZ
equation.  In 2+1 dimensions it is not completely settled whether the
large distance behaviors of KS and KPZ fall in the same universality class,
different approaches leading to conflicting
results \cite{ks2+1}.

In contrast to the KS equation, Eq.\ (\ref{eqn}) is anisotropic, and
explicitly contains a noise term. One can study Eq.\ (\ref{eqn}) using
a dynamic renormalization group (DRG) procedure.  The linear instability
generated by the negative surface tension leads to additional
singularities in
the Feynman diagrams of the perturbation expansion. This is related
to the presence of a characteristic length scale in the system,
generated by the competition between surface tension and surface
diffusion, $\ell_c = \sqrt{K/|\nu|}$, where
$\nu$ is the largest in absolute value of the negative surface
tension coefficients. Below we discuss the possible scaling
behaviour predicted by (\ref{eqn}) based on results obtained from DRG
calculations performed on its isotropic version \cite{clb}, and the results
available in the literature for other limits.
The complete scaling picture should be provided by
either a DRG analysis or a numerical integration of (\ref{eqn}).

The scaling behavior depends on the relative signs of
 $\nu_x$, $\nu_y$, $\lambda_x$ and $\lambda_y$
\cite{note2}.  The variations of these coefficients
as functions of $a_{\sigma}$
and $\theta$ lead to the phase diagram  shown  in Fig.\ \ref{fig2}.  In the
following we discuss the various scaling regimes separately.

{\it Region I ---} For small $\theta$ both $\nu_x$ and $\nu_y $ are
negative. As discussed by Bradley and Harper \cite{bh}
and experimentally studied by Chason {\em et al.} \cite{chason},
a periodic structure dominates the
surface morphology, with ripples oriented along the direction ($x$ or $y$)
which presents the largest absolute value for its surface tension
coefficient. The wavelength of the ripples is $\lambda_c = \sqrt{2}
\ell_c$.

The large length scale behavior $\ell \gg \ell_c$ is expected to be
different. Now both nonlinearities and the noise may become
relevant. A DRG analysis performed for the isotropic case of Eqn.\
(\ref{eqn}) \cite{clb,gb} suggests that inside region $I$, the
nonlinearities $\lambda_{x,y}$ flow to zero.  The surface is
logarithmically rough, with a roughness exponent $\alpha=0$.  While in
regions $Ia$-$c$ the coefficients of the nonlinear terms may change
sign as $a_{\sigma}$ or $\theta$ vary, we do
not expect this to modify the scaling behavior.

{\it Region II ---} This region is characterized by a positive $\nu_x$
and a negative $\nu_y$. Now the
periodic structure associated with the instability is directed along
the $y$ direction and is the dominant morphology at scales $\ell \sim
\ell_c$.
Much less is known about the scaling behaviour for large length scales.
Preliminary results of a DRG calculation indicate that
the surface may  be
stabilized by the positive $\nu_x$ and the nonlinearities,
and  (\ref{eqn})  reduces to the   AKPZ  equation.
If this is the case, then  the surface roughness
increases algebraically in
region $IIa$, whereas logarithmic scaling characterizes region $IIb$.

Even though several aspects of the scaling behavior predicted by
(\ref{eqn}) and (\ref{coefs2}) remain to be clarified, we believe that these
equations contain the relevant ingredients for understanding
roughening by ion bombardment \cite{last}.
To summarize, at short length
scales the morphology consists of a periodic structure oriented along
the direction determined by the largest in absolute value of the negative
surface tension coefficients \cite{chason}. Modifying the values of
$a_{\sigma}$ or $\theta$ changes the orientation of the ripples
\cite{carter,bh}.
At large length scales we expect two
different scaling regimes. One is characterized by KPZ exponents, which
might be observed in region $IIa$ in Fig.\
\ref{fig2}. Indeed, the values of the exponents reported by Eklund {\em
et al.} \cite{eklund} are consistent within the experimental errors
with the KPZ exponents in 2+1 dimensions. The other regime is characterized by
logarithmic scaling ($\alpha=0$), which has not been observed
experimentally so far. We believe that it can
be experimentally tested for suitable (e.g, small) values of the angle
of incidence. Moreover by tuning
the values of $\theta$ and/or $a_{\sigma}$ one may
induce transitions among the different scaling behaviors.  For
example, fixing $a_{\sigma}$ and increasing the
value of $\theta$ would lead from logarithmic (regions $Ia$ and $Ic$)
to KPZ scaling ($IIa$), and again logarithmic scaling ($IIb$) for
large enough angles. Also, the measurement of
the erosion velocity may help to test the applicability of
Eqn.\ (\ref{eqn}). Taking spatial and noise averages in (\ref{eqn}),
the terms contributing to the erosion velocity are the ones proportional to
$v_0$, $\xi_x$, $\xi_y$, $\lambda_x$ and $\lambda_y$. In the stationary
state, modifying $\theta$ or $a_{\sigma}$ changes the average yield,
allowing for a comparison of the experimental yield curves with the ones
predicted by (\ref{eqn}).

The experimental verification of the above possibilities would
constitute an important step to elucidate the interplay between the
mechanisms leading to the different morphologies and dynamics for
sputter-etched surfaces. It will provide as well
additional insight into the scaling behaviors to be expected from
equation (\ref{eqn}).

We would like to acknowledge discussions, comments and encouragement
by L.\ A.\ N.\ Amaral, K.\ B.\ Lauritsen, H.\ Makse and H.\ E.\ Stanley.
R.\ C.\ acknowledges a postdoctoral Fellowship of the Spanish Ministerio de
Educaci\'on y Ciencia.

\begin{figure}
\caption{Reference frames for the computation of the erosion velocity at
point $O$. Inset: Following a straigh
trajectory (solid line) the ion penetrates an average distance $a$
inside the solid (dotted line) after which it completely spreads out
its kinetic energy. The dotted curves are equal energy
contours. Energy released at point $P$ contributes to erosion at $O$.}
\label{fig1}
\end{figure}

\begin{figure}
\caption{Phase diagram for the isotropic case $\sigma=\mu=1$.
Region Ia: $\nu_x<0$, $\lambda_x <0$, $\lambda_y > 0$;
Region Ib: $\nu_x<0$, $\lambda_x >0$, $\lambda_y < 0$;
Region Ic: $\nu_x<0$, $\lambda_x >0$, $\lambda_y > 0$;
Region IIa: $\nu_x>0$, $\lambda_x >0$, $\lambda_y > 0$;
Region IIb: $\nu_x>0$, $\lambda_x >0$, $\lambda_y < 0$.}
\label{fig2}
\end{figure}



\begin{references}
\bibitem{notexi}
In the expansion of (\ref{hv}) two additional
nonlinearities are obtained on the rhs of Eqn.\ (\ref{eqn}): $\xi_x
\frac{\partial h}{\partial x} \frac{\partial^2 h}{\partial x^2} +
\xi_y \frac{\partial h}{\partial x} \frac{\partial^2 h}{\partial y^2}$,
where $\xi_x$ and $\xi_y$ are functions of $a$, $\sigma$, $\mu$ and $\theta$
\cite{us}. One can see that $\xi_x$ and $\xi_y$ are irrelevant at large length
scales
using the known values of $\alpha$ and $z$ \cite{numer} for both the nonlinear
and the
linear fixed points in 2+1 dimensions.
\bibitem{note1}
The $\theta=0$ case has to be treated separately and is discussed in Ref.
\cite{us}.
\bibitem{note2}
The terms $-v_0$ and $\partial_x h$ can be reabsorbed by
a change of variables to a comoving frame, and do not
affect the scaling properties.
\bibitem{last}
Our analysis describes the roughening process in the small slope approximation.
However, at late stages additional non-linear effects, such as shadowing
\cite{bales}, may become relevant.
\end{references}
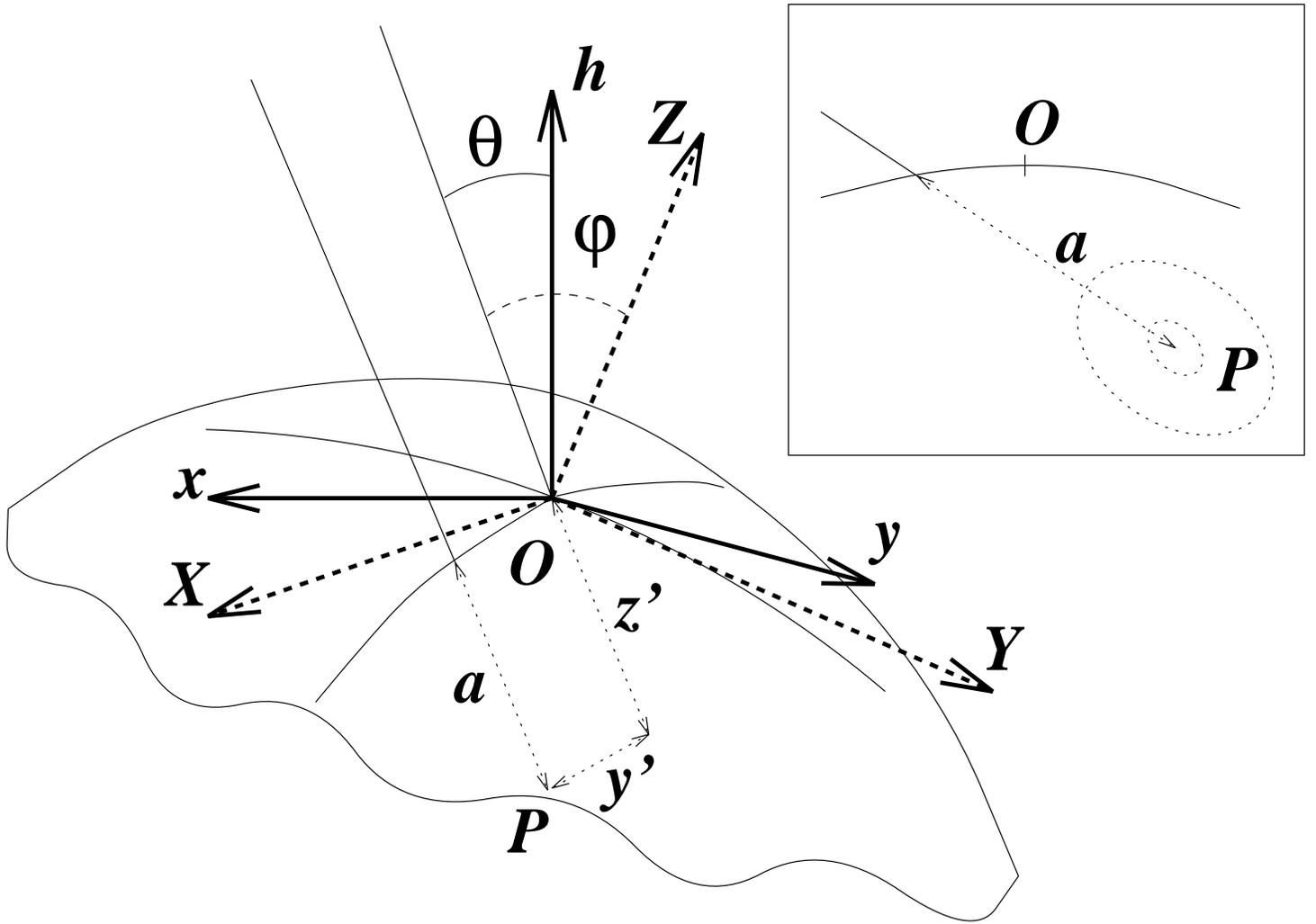
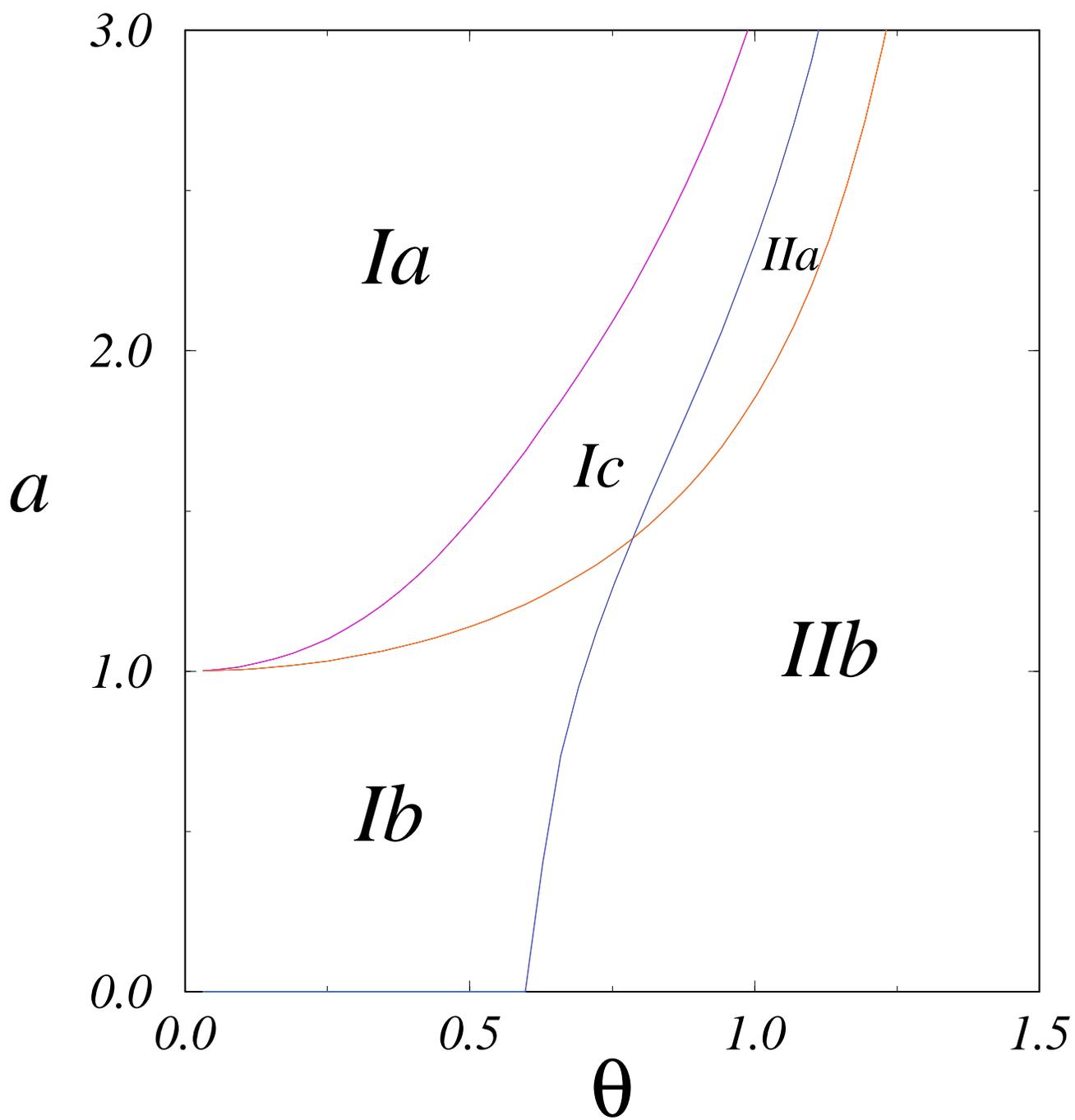
\end{document}